\documentclass[aps,prl,twocolumn,showpacs,superscriptaddress,footinbib]{revtex4-2} 
\usepackage{amssymb}
\usepackage{amsmath}
\usepackage[colorlinks=true,linkcolor=blue,citecolor=blue, urlcolor=blue]{hyperref} 
\usepackage{tikz}
\usepackage{graphicx}
\usepackage[utf8]{inputenc}
\definecolor{uibred}{RGB}{167, 38, 47}

\makeatletter
\newcommand{\fmarkiii}{*}
\newcommand{\fmarki}{\ensuremath{\dagger}}
\newcommand{\fmarkii}{\ensuremath{\ddagger}}
\def\@fnsymbol#1{{\ifcase#1\or \fmarki\or \fmarkii\or \fmarkiii \else\@ctrerr\fi}}
\makeatother

\newcommand{\vp}{\mathbf{p}}
\newcommand{\vk}{\mathbf{k}}
\newcommand{\mysection}[1]{{\vspace{10 pt}\noindent \emph{{#1}.--}}}

\begin{document}

\title{Prescaling relaxation to nonthermal attractors}

\author{Michal P.\ Heller} \email{michal.p.heller@ugent.be}
\affiliation{Department of Physics and Astronomy, Ghent University, 9000 Ghent, Belgium}

\author{Aleksas Mazeliauskas} \email{a.mazeliauskas@thphys.uni-heidelberg.de}
\affiliation{Institut f\"{u}r Theoretische Physik, Universit\"{a}t Heidelberg, 69120 Heidelberg, Germany}

\author{Thimo Preis} \email{preis@thphys.uni-heidelberg.de}
\affiliation{Institut f\"{u}r Theoretische Physik, Universit\"{a}t Heidelberg, 69120 Heidelberg, Germany}

\begin{abstract}
We study how isotropic and homogeneous far-from-equilibrium quantum systems relax to nonthermal attractors, which are of interest for cold atoms and nuclear collisions. We demonstrate that a first-order ordinary differential equation governs the self-similar approach to nonthermal attractors, i.e., the prescaling. We also show that certain natural scaling-breaking terms induce logarithmically slow corrections that prevent the scaling exponents from reaching the constant values during the system's lifetime. We propose that, analogously to hydrodynamic attractors, the appropriate mathematical structure to describe such dynamics is the transseries. We verify our analytic predictions with state-of-the-art $2$PI simulations of the large-$N$ vector model and QCD kinetic theory.
\end{abstract}

\maketitle

\mysection{Introduction} 
Thermalization of isolated quantum many-body systems is an important contemporary research problem of a broad scope. Its relevance ranges from cold atom systems, through QCD in ultrarelativistic nuclear collisions all the way to gravity and black hole physics~\cite{Berges:2020fwq}. 
Given the complexity of modeling quantum many-body dynamics and the richness of non-equilibrium phenomena, emergent regularities that form a basis for a quantitative understanding are of particular interest.

In this work we are concerned with an important instance of such an emergent regularity: far-from-equilibrium self-similar time evolution of nonthermal attractors, also known as nonthermal fixed points. These phenomena are transient stages in the thermalization dynamics, whose defining feature is self-similar scaling behavior in time.  Consider a momentum distribution function $f(t,\vp)$ of a homogeneous and isotropic system, where $t$ is time and $\vp$ spatial momentum. The system reaches a nonthermal attractor, when $f$ scales with time in a characteristic momentum range
\begin{equation}
\label{eq.scaling_iso}
    f(t,\vp) = (t/t_{\mathrm{ref}})^{\alpha_\infty} f_S((t/t_{\mathrm{ref}})^{\beta_\infty} |\vp|)
\end{equation}
with \emph{constant} scaling exponents $\alpha_\infty$ and $\beta_\infty$. Indeed, such behavior corresponds to a vast reduction in the complexity, as the knowledge of the distribution function at some time allows one to determine the distribution function at a different time by a simple rescaling. 

Nonthermal attractors appear in the studies of isolated quantum systems across a wide range of energy scales:  ultracold quantum gases~\cite{Prufer:2018hto,Erne:2018gmz,Glidden:2020qmu,
Navon:2016em,johnstone2019evolution,Helmrich:2020sgn,Garcia-Orozco:2021hkx,Martirosyan:2023mml,Huh:2023xso,Lannig:2023fzf}, ultrarelativistic nuclear collisions~\cite{Baier:2000sb,Berges:2013eia,Kurkela:2015qoa} and early universe cosmology~\cite{Micha:2002ey,Berges:2008wm}. Despite significant interest in nonthermal attractors, a quantitative understanding of how a system approaches a nonthermal fixed point remains elusive~\cite{Mazeliauskas:2018yef,Mikheev:2018adp,Brewer:2022vkq,Mikheev:2022fdl,Preis:2022uqs}.

In~\cite{Mazeliauskas:2018yef} it was proposed that even prior to reaching the nonthermal attractor~\eqref{eq.scaling_iso} the system can exhibit prescaling, where $f$ has already assumed the fixed-point shape $f_S$ but continues to evolve with time dependent scaling exponents $\alpha(t)$ and $\beta(t)$
\begin{equation}
\label{eq.nonexp_prescaling}
    f(t,\vp) =A(t)\, f_S(B(t)|\vp|).
\end{equation}
The prescaling factor $A(t)=\exp[\int_{t_0}^t dt^\prime \alpha(t^\prime)/t^\prime]$ reduces to the fixed-point scaling of Eq.~(\ref{eq.scaling_iso}) when $\alpha(t)$ approaches~$\alpha_{\infty}$. The same holds for $B(t)$ in terms of $\beta(t)$. 

Given that scaling~\eqref{eq.scaling_iso} is an asymptotic late time statement known to be reached slowly, the systems of interest might in fact spend a much greater fraction of their lifetime prescaling~\eqref{eq.nonexp_prescaling} rather than scaling~\eqref{eq.scaling_iso}. Therefore, a quantitative understanding of prescaling is as important as understanding scaling itself.

In our work, we develop a simple theoretical description of prescaling dynamics that uses the same assumptions as the ones used to derive scaling. We test our predictions using strongly-correlated large-$N$ vector model and weak coupling
QCD kinetic theory simulations.

\mysection{Scaling implies prescaling} 
Understanding prescaling requires identifying laws governing time evolution of $A(t)$ and $B(t)$ (or, alternatively, $\alpha(t)$ and $\beta(t)$). As we show, these laws have a surprisingly simple origin and form.  

The key role in deriving scaling~\eqref{eq.scaling_iso} is played by conserved quantities: particle number density $n = \int d^d\vp\, f/(2\pi)^d$ or energy density $ \mathcal{E} = \int d^d\vp\, \omega_\vp f/(2\pi)^d$, where~$d$ is the number of spatial dimensions and $\omega_\vp$ is the dispersion relation of particles. We focus on~$\omega_\vp \sim |\vp|^z$.
Applying conservation of $n$ or $\mathcal{E}$ in the momentum regime of interest imposes the relation $\alpha_\infty = \sigma \, \beta_\infty$ between the scaling exponents~\cite{PineiroOrioli:2015cpb}. When $n=\text{const}$, then $\sigma = d$, while $\mathcal{E}=\text{const}$ gives $\sigma = (d+z)$. The conserved quantities are local in time, which means that they in fact constrain also prescaling exponents in exactly the same way: $\alpha(t) = \sigma \, \beta(t)$. Equivalently, $A(t) = B(t)^{\sigma}$. This implies that there is only one independent degree of freedom in the isotropic and homogeneous prescaling, which we will choose to be~$B(t)$.

The time evolution for the independent prescaling factor $B(t)$ is still subject to the equation of motion for~$f$. In the case of a kinetic theory it is given by the Boltzmann equation with collision kernel~${\cal C}[f]$
\begin{equation}
\label{eq.nonexp_Boltzmann}
    \partial_t f(t,\vp) = {\cal C}[f](t,\vp). 
 \end{equation}

In the present section, we assume the collision kernel to be a homogeneous functional of particles momenta, i.e., to simply scale under Eq.~(\ref{eq.nonexp_prescaling}) by 
$A(t)^{\mu_{\alpha}} B(t)^{\mu_{\beta}} \equiv B(t)^{\sigma \mu_\alpha+\mu_\beta}$ for some real numbers $\mu_{\alpha,\beta}$.
This assumption applies to many (but not all) collision kernels describing nonthermal attractors (see the Supplemental Material~\cite{supp} for explicit examples). Typically overoccupation singles out terms with the highest power of the distribution function and associated matrix elements often happen to scale homogeneously under rescalings of momenta. For such collision kernels, we can separate time-dependent and (rescaled) momentum-dependent contributions by substituting the prescaling ansatz \eqref{eq.nonexp_prescaling} in the Boltzmann equation \eqref{eq.nonexp_Boltzmann} and using $A(t)=B(t)^\sigma$,
\begin{equation}
  \frac{B(t)^{1-1/\beta_\infty}}{\partial_t B(t)}=\frac{1}{D_1}=\frac{\left[ \sigma + \bar{\vp}\cdot \partial_{\bar{\vp}}\right] f_S(\bar{\vp})}{{\cal C}[f_S](\bar{\vp})}\label{eq.nonexp_Bode_general},  
\end{equation}
where $\bar{\vp}=B(t)\,\vp$ is the rescaled momentum, $1/\beta_\infty=(1-\mu_\alpha)\sigma -\mu_\beta$, and $D_1$ is a separation of variables constant. The intrinsic time dependence of our setup implies nonzero~$D_1$, which can be fixed at any time in the prescaling evolution and we choose $D_1=\beta(t_0)/t_0$.

The idea of separation of variables in the context of nonthermal attractors appeared already in~\cite{Micha:2004bv}, but for homogeneous $\mathcal{C}[f]$ only solutions with constant scaling exponents were considered. Our key observation here is that prescaling is encapsulated by the general solution
\begin{subequations}
\label{eq.nonexp_general_sol}
    \begin{align}
 B(t) &= \left(\tfrac{t-t_{*}}{t_{\mathrm{ref}}}\right)^{\beta_{\infty}}\label{eq.nonexp_B_general} \approx \left( \tfrac{t}{t_{\mathrm{ref}}} \right)^{\beta_{\infty}} \left( 1 -\beta_{\infty} \tfrac{t_*}{t} + \ldots \right) \\
        \beta(t) &= \beta_\infty \tfrac{t}{t-t_{*}} \approx \beta_\infty + \beta_{\infty} \tfrac{t_*}{t} + \ldots\quad . \label{eq.nonexp_beta_general}
    \end{align}
\end{subequations}
From Eq.~\eqref{eq.nonexp_general_sol} it is clear that prescaling induces power-law corrections to scaling. The prescaling originates from the presence of nonzero $t_{*}=t_0(1-\beta_\infty/\beta_0)$, where $t_0$ and $\beta_0\equiv \beta(t_0)$ correspond to the initial data. Its appearance comes as no surprise: the dynamics of the system in question is time translationally invariant and therefore $t$ appearing in formulas needs to be measured with respect to some time, $t_{*}$. The reason why it does not appear in Eq.~\eqref{eq.scaling_iso} is because the exact scaling is an asymptotic late time statement and dependence on $t_*$ drops. Note that while $\beta_{\infty}$ and $\alpha_{\infty}$ are theory specific and independent of initial conditions, $t_*$ will depend on a chosen initial state.

Before we move to testing Eq.~\eqref{eq.nonexp_general_sol} using ab initio solutions of quantum dynamics, let us reiterate that this result originates from the Boltzmann equation and pertinent conservation laws. These are exactly the same constraints as used in a conventional scaling analysis~\cite{PineiroOrioli:2015cpb}. Prescaling in the case of collision kernels being homogeneous functionals of momenta can therefore be understood as a direct consequence of the existence of scaling.

\begin{figure}[t!]
    \centering
        \includegraphics[width=1\columnwidth]{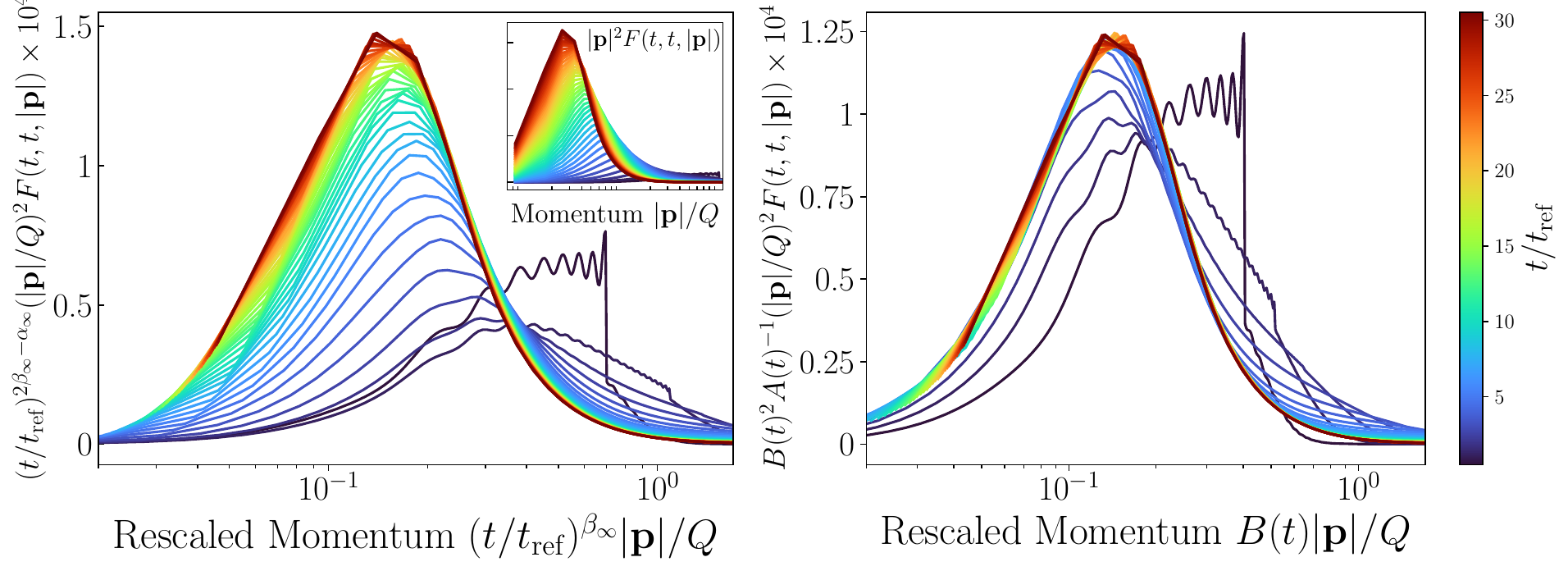}
        \includegraphics[width=0.93\columnwidth]{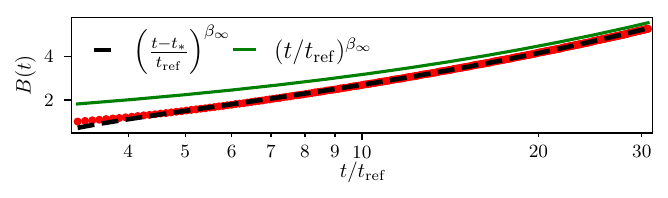}    \caption{(Top) Statistical function $|\vp|^2 F(t,t,|\vp|)$ from 2PI simulations of large-$N$ vector model rescaled with (left) fixed point exponents $(\alpha_\infty,\beta_\infty)=(3/2,1/2)$ and (right) with prescaling factors $A(t),B(t)$, where $t_{\mathrm{ref}}\, Q\sim 93$. The unrescaled distribution is shown in the inset. (Bottom) Extracted $B(t)$ compared with Eq.~\eqref{eq.nonexp_B_general} and asymptotic form.
    }
    \label{fig.nonexp_F2piresc}
\end{figure}
\mysection{Prescaling in large-$N$ vector model}
We begin by testing Eq.~(\ref{eq.nonexp_general_sol}) against the full quantum dynamics of a next-to-leading order (NLO) large-$N$ vector model at small coupling $\lambda$. This model features a dual cascade~\cite{PineiroOrioli:2015cpb,Berges:2016nru} consisting of an inverse particle cascade in the IR and a direct energy cascade to the UV. We focus on the IR scaling, which is characterized by effective particle number conservation with $(\alpha_\infty,\beta_\infty)=(d/2,1/2$) and was also realized in cold quantum gases~\cite{Prufer:2018hto,Glidden:2020qmu}. This fixed point arises for large occupations $f\sim 1/\lambda \gg 1$ such that its description requires going beyond a standard kinetic theory analysis. Large-$N$ kinetic theory addresses this regime due to inclusion of relevant resummations~\cite{PineiroOrioli:2015cpb,Walz:2017ffj}. The corresponding collision kernel scales homogeneously with $\mu_\alpha=1$ and $\mu_\beta=-2$~\cite{PineiroOrioli:2015cpb}.

We perform ab initio studies of this fixed point in $3+1$ dimensions using the $2$PI formalism following~\cite{Preis:2022uqs} to validate Eq.~\eqref{eq.nonexp_general_sol} and the underlying assumptions. Below the mass gap the equal-time statistical function $F(t,t,|\mathbf{p}|)$ reduces to $f(t,|\mathbf{p}|)$~\cite{PineiroOrioli:2015cpb}, where the system is initialized with $f(t_0,|\mathbf{p}|)=f_0\theta(Q-|\mathbf{p}|)$ with $f_0=100/\lambda$, $\lambda=0.01$ and $Q$ is the characteristic hard scale far from equilibrium. Following~\cite{Mazeliauskas:2018yef} we extract the prescaling factors $A(t)$ and $B(t)$ from the time evolution of integral moments of $F(t,t,|\mathbf{p}|)$~\cite{supp}, e.g., $B(t) = n(t)\mathcal{E}(t_0)/(\mathcal{E}(t)n(t_0))$~\footnote{In the extraction of $B(t)$ for the IR attractor in Fig.~\ref{fig.nonexp_F2piresc} of the main text we used the hard scale $Q(t)$ as a cutoff in the computation of the integral moments. $Q(t)$ separates the IR and UV scaling regions of the corresponding dual cascade and we dynamically determine it from the minimum at intermediate momentum scales of the function $|\vp|^2 f(t,|\vp|)$.}. In the upper right panel of Fig.~\ref{fig.nonexp_F2piresc} we show how rescaling with $A(t)$ and $B(t)$ leads to an early collapse of distributions at different times, while a considerable spread remains when rescaling with the fixed point exponents (left panel). The evolution of the extracted $B(t)$ is shown in the lower panel to be well described by Eq.~(\ref{eq.nonexp_B_general}) already at early times (dashed black line), and only asymptotes to the corresponding fixed point scaling behavior~(\ref{eq.scaling_iso}) (solid green line). The full NLO quantum dynamics are shown to be captured remarkably well by our effective kinetic description of Eq.~\eqref{eq.nonexp_general_sol} already from times close to initialization.

\mysection{Prescaling in isotropic QCD kinetic theory}
We move now to studying prescaling dynamics in QCD, whose nonthermal fixed point plays an important role in our understanding of thermalization dynamics in weakly-coupled models of ultrarelativistic nuclear collisions~\cite{Berges:2020fwq}. We use QCD kinetic theory, where the evolution of the color and polarization averaged gluon distribution function $f(t,\vp)$ is described by $2\leftrightarrow 2$ and $1\leftrightarrow 2$ processes~\cite{Arnold:2002zm}: $\partial_t f(t,\vp) = {\cal C}^{2\leftrightarrow 2}[f](t,\vp)+{\cal C}^{1\leftrightarrow 2}[f](t,\vp)$, see~\cite{supp,Arnold:2002zm,Keegan:2016cpi,Kurkela:2018oqw}. 
Nonthermal fixed points can be reached from a wide range of initial conditions including large occupation numbers~\cite{Berges:2014bba,Deng:2018xsk,Chantesana:2018qsb,PineiroOrioli:2018hst,Schmied:2018upn,Boguslavski:2019ecc,Boguslavski:2021kdd}, which we implement via
$
f(t_i,\vp) = n_0/g^2  \exp\left[-\vp^2/Q^2 \right].
$
Here $g^2$ is the square of the coupling and $n_0$ is the initial occupation. We consider $n_0=1$ and $g^2=10^{-8}$. To obtain the precise late time behavior we initialize at $t_iQ=0$ and evolve for very long times until $t_fQ=10^{8}$.  Results will be given in units of the characteristic energy scale $Q$ as given by the maximum of $|\vp|^2 f(t_i,\vp)$. We discuss explicitly only the pure gluon simulations where the scaling phenomenon is encountered after checking that our results do not change under the inclusion of quark/anti-quark dynamics.

A scaling analysis for the vacuum QCD collision kernel together with energy conservation $\sigma=d+z=4$ reveals the direct energy cascade fixed point $(\alpha_\infty,\beta_\infty)=(-4/7,-1/7)$, see \cite{Schlichting:2012es,Berges:2013fga,AbraaoYork:2014hbk}. However, the overall scaling of the elastic collision kernel is broken by the presence of the Debye mass $m_D(t)^2\sim \int d^3 \vp f(t,\vp)/p \sim A(t) B(t)^{-2} \bar{m}^2_D$ that regulates soft elastic scatterings where $\bar{m}^2_D\equiv m_D(t_0)^2$~\cite{supp}. With $\alpha_\infty,\beta_\infty <0$, the Debye mass decreases over time such that the violation of scaling only leads to a delay in the approach to the fixed point.
\begin{figure}[t!]
    \centering
    \includegraphics[width=0.98\columnwidth]{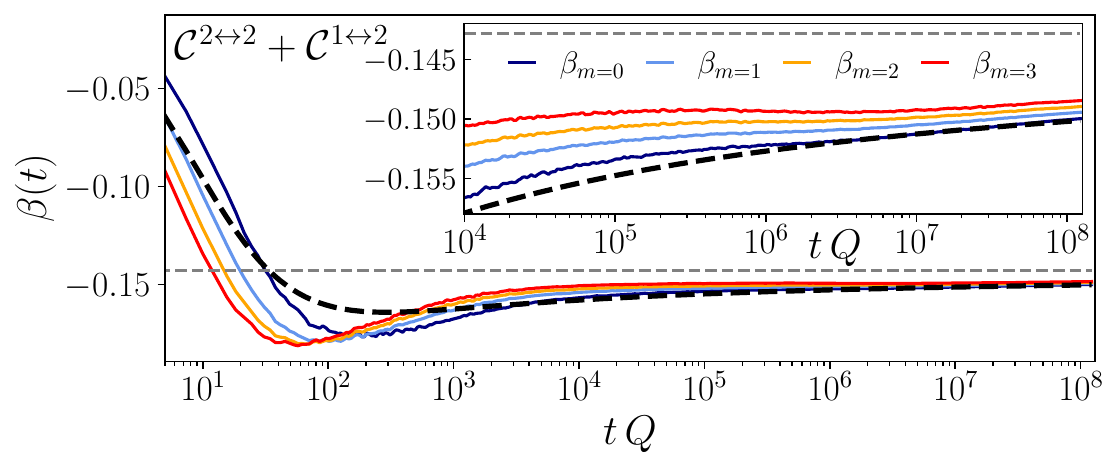}    \caption{Comparison between prescaling exponents extracted from QCD kinetic theory simulations with scaling-breaking elastic processes and the analytical prescaling expectation of Eq.~\eqref{eq.nonexp_betaode_FP} (dashed black line). Displayed are $\beta_{m=0,1,2,3}(t)$ computed from different moments of the distribution function~\cite{supp}.}
    \label{fig.nonexp_prescaling_FPpred}
\end{figure}
This (diminishing) breaking of scaling for QCD kinetic theory is demonstrated in Fig.~\ref{fig.nonexp_prescaling_FPpred}, where we extract $\beta(t)$ (solid color lines) from different moments of the distribution function~\cite{supp}. We do not display $\alpha(t)$ explicitly as we find the scaling relation $\alpha(t)/\beta(t)=4$ realized to a very good accuracy. $\beta(t)$ is observed to approach the fixed point value $\beta_\infty=-1/7$ (dashed gray line) but a finite deviation remains due to $m_D(t)$ even after eight orders of magnitude in simulation time. The decreasing but finite spread between different moments further demonstrates that different momenta of the distribution function approach the fixed point on different timescales. A fast convergence of the moments and collapse to $\beta_\infty$ is in contrast found for simulations without the elastic collision kernel $\mathcal{C}^{2\leftrightarrow 2}$~\cite{supp}. In the following we will study these deviations analytically in a small-angle scattering approximation of $\mathcal{C}^{2\leftrightarrow 2}$.

\mysection{Effect of scaling breaking terms in the Fokker-Planck approximation}
The breaking of scaling inhibits prescaling exponents extracted from different moments to share the same universal prescaling dynamics. Nevertheless, at qualitative level the scaling dynamics can be reasonably modeled via the Fokker-Planck (FP) approximation~\cite{Mueller:1999pi,Blaizot:2013lga,Schlichting:2019abc}. This approach assumes the dominance of small angle scatterings and has previously been used in the context of nonthermal attractors~\cite{Tanji:2017suk} and prescaling~\cite{Brewer:2022vkq,Mikheev:2022fdl}.
We will compare our analytical results from the FP approximation to simulations using the full QCD collision kernel. The corresponding FP collision kernel allows us to factorize the 
scaling-breaking Coulomb logarithm, which involves the ratio of the UV scale, the characteristic gluon energy $\langle p\rangle$, and the IR scale, the Debye mass $m_D$
\begin{align}
\label{eq.fp_collision}
    {\cal C}^{\mathrm{FP}}[f](t,\vp)
    &=\frac{A(t)^3}{B(t)} \log\left[\frac{\langle \bar{p}\rangle}{A(t)^{\frac{1}{2}} \bar{m}_D} \right]\tilde{C}^{\mathrm{FP}}[f_S](\bar{\vp}),
\end{align}
where we only display terms relevant for the scaling analysis~\cite{supp}.
We note that the time-dependence due to the Coulomb log can be identified with the background term in the formalism of~\cite{Micha:2004bv}. The FP-kernel does not scale homogeneously and the solution ~\eqref{eq.nonexp_general_sol} does not apply. However, separation of variables can still be performed as the necessary property is factorization in $t$ and $\bar{\vp}$-dependencies. Upon separating variables with associated constant $D_2$, see Eq.~(\ref{eq.nonexp_Bode_general}), and relating $\alpha(t)=\sigma\beta(t)$ as before, we now obtain 
\begin{equation}
\label{eq.nonexp_Bode_FP}
    \partial_t B(t) = \frac{\beta_{0}}{t_{0}} B(t)^{1-1/\beta_\infty} \left[1-\frac{\sigma}{2}\frac{\log[B(t)]}{\log[\langle\bar{p}\rangle/\bar{m}_D]}\right].
\end{equation}
Equation~\eqref{eq.nonexp_Bode_FP} can be directly integrated, but then $B(t)$ appears as an argument of a nontrivial transcendental function. A more useful approach is to derive from Eq.~\eqref{eq.nonexp_Bode_FP} a~second order differential equation for $\beta(t)$
\begin{equation}
\label{eq.nonexp_betaode_FP}
    \frac{\ddot{\beta}(t)}{\beta(t)} =\frac{\dot{\beta}(t)^2}{\beta(t)^2}+\frac{14 \dot{\beta}}{t}- \frac{(7\beta(t)+1)^2}{t^2},
\end{equation}
which is directly solved by $\beta(t)=\beta_\infty=-1/7$. The solution to Eq.~\eqref{eq.nonexp_betaode_FP} (dashed black line) is shown in Fig.~\ref{fig.nonexp_prescaling_FPpred} to capture well the evolution of $\beta(t)$ obtained from solving the full QCD kinetic theory collision kernel from times shortly after initialization of the system over more than eight orders of magnitude. We obtain this result by solving Eq.~(\ref{eq.nonexp_betaode_FP}) with initial conditions for $\dot{\beta}_{0}$ determined consistently from the full QCD kinetic theory data at $t_0$~\cite{supp}. In the inset we show that solving Eq.~(\ref{eq.nonexp_betaode_FP}) shortly after initialization describes qualitatively well the late-time dynamics.

With the applicability of Eq.~\eqref{eq.nonexp_betaode_FP} demonstrated in Fig. \ref{fig.nonexp_prescaling_FPpred}, we now use it to study the prescaling dynamics in the vicinity of the fixed point. We can linearize Eq.~(\ref{eq.nonexp_betaode_FP}) in perturbations $\delta \beta(t)$ around the fixed point value $\beta_\infty=-1/7$, which yields a power-law decay from below $\delta \beta(t) \sim -1/t$. The full solution to Eq.~\eqref{eq.nonexp_betaode_FP} at late times however turns out to be governed by logarithmic corrections not captured by this linearization procedure. We find that a consistent late-time solution to Eq.~\eqref{eq.nonexp_betaode_FP} is given by
\begin{subequations}
\begin{eqnarray}
\hspace{-18 pt}    \beta(t) &\approx& \beta_{\infty} + \sum_{m=1}^{\infty} \sum_{n =0}^{m-1} \beta_{m, n} \frac{\log(\log(t \, Q))^{n}}{\log(t \, Q)^m} +\mathcal{O}\left(\tfrac{1}{t}\right)\label{eq.lateansatz}\\
    &\approx& \beta_\infty \left[ 1+ \frac{1}{\log(t\, Q)}\right],\label{eq.nonexp_betaode_FP_AsLogSol}
\end{eqnarray}
\end{subequations}
where we used $Q$ as the reference scale but emphasize that the choice of a constant does not matter at late enough times. Similar late-time power-law~\cite{Mikheev:2022fdl} corrections from linearization and late-time logarithmic~\cite{Brewer:2022vkq} corrections induced by the temporal evolution of the Coulomb logarithm were found in the FP approximation for the Baier-Mueller-Schiff-Son~\cite{Baier:2000sb} fixed point in longitudinal expanding plasma.

 The simple power-law approach to the fixed point found in the absence of scaling breaking terms in Eq.~(\ref{eq.nonexp_beta_general}) is therefore enriched to involve both fast (power-law) and slow (inverse powers of logarithms and slower) behavior. This discussion is reminiscent of the transseries form~\cite{Dorigoni:2014hea,Aniceto:2018bis} for late time dynamics of the energy-momentum tensor of matter undergoing longitudinal boost-invariant expansion~\cite{Heller:2015dha,Basar:2015ava,Aniceto:2015mto,Aniceto:2018uik}. There slow modes came from relativistic hydrodynamics and exponentially faster modes from transient excitations. Here slow modes come from the Debye mass breaking the homogeneity of the collision kernel with respect to rescalings of momenta and fast modes are the original prescaling excitations encountered already in~\eqref{eq.nonexp_beta_general}. Similarly to~\cite{Heller:2015dha,Basar:2015ava,Aniceto:2015mto,Aniceto:2018uik}, it is not difficult to gather finite order indications that the series containing slow modes~\eqref{eq.lateansatz} has a vanishing radius of convergence with $\beta_{m+1,0}/\beta_{m,0} \sim m$~\cite{supp}. Curiously, the leading (at each $m$) doubly logarithmic term behaves geometrically: $\beta_{m+1,m}/\beta_{m,m-1} = -1$. It would be interesting to develop systematic understanding of this behavior, including resummations of the resulting transseries. A~good starting point might be analysis of the Painlev{\'e} I equation in~\cite{Garoufalidis:2010ya,Aniceto:2011nu}, which is also second order and exhibits expansion in three building blocks analogous to ours $t$, $\log(t\,Q)$ and $\log{(\log{(t\, Q)})}$.

\begin{figure}[t!]
    \centering
     \includegraphics[width=1\columnwidth]{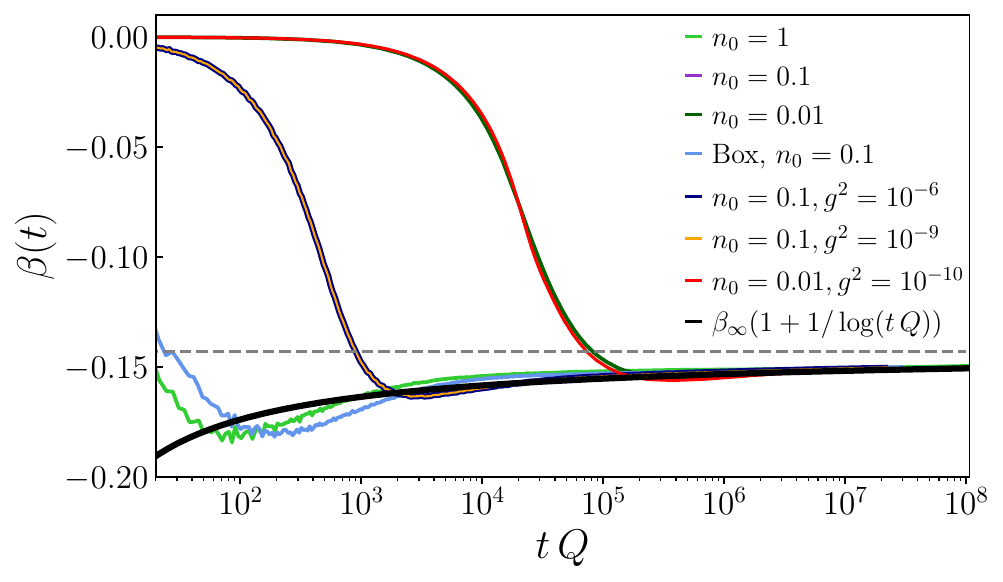}    \caption{Comparison of prescaling trajectories in QCD kinetic theory simulations. Solid lines correspond to prescaling exponents extracted from simulations with different initial conditions, where for visibility only results from the first moment are displayed. 
    }
    \label{fig.nonexp_FP_attractor}
\end{figure}
In Fig.~\ref{fig.nonexp_FP_attractor} we visualize the attractive nature of the prescaling dynamics by extracting prescaling exponents from different initial conditions. All simulations are initialized with variations in parameters of the class of initial condition used in this work apart from the data represented by light blue, which uses box initial conditions $f(t_i,\vp)=n_0/g^2 \theta(Q-|\vp|)$. The prescaling exponents extracted from different simulations are all found to converge to a universal late-time behavior, which we additionally show is well described by Eq.~(\ref{eq.nonexp_betaode_FP_AsLogSol}) (solid black line). Furthermore, we want to emphasize the similarity between the behavior shown in Fig.~\ref{fig.nonexp_FP_attractor} and hydrodynamic attractors, where different solutions converge to a single universal curve which at sufficiently late times is described by relativistic hydrodynamics~\cite{Heller:2015dha,Romatschke:2017vte,Jankowski:2023fdz}.

The above analysis has an important bearing on the appearance of scaling. The regime when the highest order terms in the collision kernel dominate parametrically ends when the typical occupancy becomes of $\mathcal{O}(1)$. This is realized if $t_f Q \geq \alpha^{-7/4}_S$~\cite{Berges:2013eia}. At that time, we have a deviation of $\delta \beta(t)/\beta \sim 1/\log( \alpha^{-7/4}_S) \lesssim 0.03$ with $g^2=10^{-8}$. As a consequence of this,  QCD kinetic theory will therefore still show percent deviations from the fixed point values when the direct energy cascade ceases and ultraviolet modes $|\mathbf{p}|/Q\geq 1$ start to thermalize.

\mysection{Conclusions}
We studied the approach of isotropic and spatially homogeneous quantum many-body systems to nonthermal attractors. 
Our results demonstrate that the prescaling is governed by a simple first-order ordinary differential equation obtained from the underlying dynamics via emergent conservation laws.

Our analytical prediction implies that prescaling entails infinitely many power-law corrections to constant scaling exponents. They conspire to a simple time off-set in the fixed point scaling.
We have successfully tested our simple formula for prescaling against ab initio simulations of a relativistic vector model QFT using 2PI formalism and QCD kinetic theory simulations. Our QCD kinetic theory simulations span eight orders of magnitude in time and provide the most accurate extraction of scaling exponents to date.

The exact scaling associated with nonthermal attractors requires the collision term to be a homogeneous functional of particle momenta at large occupations. For QCD kinetic theory this property is violated by the Debye mass term that regulates the Coulomb divergence in the elastic scattering matrix element. We demonstrate that exact scaling exponents are not reached during the lifetime of the system.
Using the Fokker-Planck approximation to QCD kinetic theory we show that the scaling-breaking Couloumb logarithm significantly enriches the prescaling dynamics. The late-time behavior is given by a factorial divergent series that includes inverse powers of logarithms and positive powers of double logarithms of time. This constitutes a striking structural similarity with theoretical descriptions of hydrodynamic attractors in the boost-invariant models of nuclear collisions.

Our work shows that prescaling is an unavoidable consequence of nonthermal attractors. Therefore our analytical predictions for prescaling can be verified experimentally in cold atom systems. Furthermore, we uncovered that scaling breaking terms generate rich prescaling dynamics that bares similarities to transseries in the context of hydrodynamic attractors. It would be fascinating to utilize the enormous degree of control in cold atom systems to induce scaling breaking terms and experimentally discover the phenomenology of transseries.

\emph{Note added:} Several months after the arXiv submission of the present work, Ref.~\cite{Gazo:2023exc} appeared and reported the observation of the prescaling solutions that we derived in Eq.~\eqref{eq.nonexp_general_sol} in an ultra-cold atom experiment.

\begin{acknowledgments}
\mysection{Acknowledgments}
We thank 
I.~Aniceto,
J.~Berges,
K.~Boguslavski,
J.~Brewer,
M.~Gałka,
A.~Kurkela,
A.~Mikheev,
J.~Noronha,
B.~Scheihing-Hitschfeld,
S.~Schlichting,
A.~Serantes,
R.~Venugopalan,
and 
Y.~Yin 
for useful discussions and comments on the draft.
The authors acknowledge support by the state of Baden-Württemberg through bwHPC and the German Research Foundation (DFG) through grant no 
INST 40/575-1 FUGG (JUSTUS 2 cluster), the DFG under the Collaborative Research Center SFB 1225 ISOQUANT (Project-ID 273811115), the Simons Foundation (Award number 994318, Simons Collaboration on Confinement and QCD Strings) and the Heidelberg STRUCTURES Excellence Cluster under Germany's Excellence Strategy EXC2181/1-390900948. 
The work of AM is funded by DFG – Project number 496831614. This project has received funding from the European Research Council (ERC) under the European Union’s Horizon 2020 research and innovation programme (grant number: 101089093 / project acronym: High-TheQ). Views and opinions expressed are however those of the authors only and do not necessarily reflect those of the European Union or the European Research Council. Neither the European Union nor the granting authority can be held responsible for them. We would like to thank KITP for its hospitality during the program ``The Many Faces of Relativistic Fluid Dynamics" supported by the National Science Foundation under Grant No. NSF PHY-1748958.
\end{acknowledgments}

\bibliography{master.bib}

\begin{appendix}
\widetext 
\renewcommand{\theequation}{A\arabic{equation}}
\renewcommand{\thefigure}{A\arabic{figure}}
\setcounter{equation}{0}
\setcounter{figure}{0}
\makeatletter
\section{Appendix}
\section{Extraction of scaling exponents}
Under the prescaling ansatz for an isotropic system, Eq.~\eqref{eq.nonexp_prescaling} in the main text, we introduce moments of the distribution function
\begin{align}
    n_m(t) &= \int \frac{d^d \vp}{(2\pi)^d} |\vp|^m f(t,\vp) = A(t) B(t)^{-(d+m)} \bar{n}_m,
\end{align}
whose evolutions are given by the dynamics of prescaling exponents \cite{Mazeliauskas:2018yef}
\begin{equation}
 \frac{d}{d\log t} \log(n_m(t)) =\alpha(t)-(d+m)\beta(t).
\end{equation}
 We can thus extract prescaling exponents from the evolution of the moments, for example,
\begin{equation}
\label{eq.nonexp_exponents_moments}
        \alpha_m(t) =(m+d)(m+1+d) \frac{d}{d\log t} \log \left[ \frac{n_m(t)^{\frac{1}{m+d}}}{n_{m+1}(t)^{\frac{1}{m+1+d}}}\right], \qquad \beta_m(t) = \frac{d}{d\log t} \log \left(\frac{n_m(t)}{n_{m+1}(t)} \right).
\end{equation}
We can similarly obtain $A(t)$ and $B(t)$ from moments of the distribution function (or equivalently from moments of the equal-time statistical function $F(t,t,|\mathbf{p}|)$ for the discussion of the infrared fixed point) according to
\begin{equation}
\label{eq.nonexp_AB_moments}
A(t) =\left[\frac{n_m(t)^{1/(d+m)}}{n_{m+1}(t)^{1/(d+m+1)}} \frac{\bar{n}^{1/(d+m+1)}_{m+1}}{\bar{n}^{1/(d+m)}_m} \right]^{(d+m)(d+m+1)}\hspace{-2cm},\hspace{2.5cm} B(t) = \frac{n_m(t)}{n_{m+1}(t)} \frac{\bar{n}_{m+1}}{\bar{n}_m},
\end{equation}
where $\bar{n}_m\equiv n_m(t_0)$ such that $B(t_0)=1$.


\section{QCD kinetic theory and prescaling}
In this work we study QCD kinetic theory which includes $2\leftrightarrow 2$ and $1\leftrightarrow 2$ collinear scattering terms. We will give the corresponding equations for the gluon sector here and refer to \cite{Kurkela:2018oqw,Berges:2020fwq} for the complete expressions including other particle species. The gluon collision kernels are parametrized as
\begin{subequations}
\label{eq.app_ekt_collisionkernels}
    \begin{align}
        \mathcal{C}^{2\leftrightarrow 2}[f](t,\vp) &= -\frac{1}{8p(N^2_c-1)} \int \frac{d^3\vk d^3 \vp^\prime d^3 \vk^\prime}{(2\pi)^9 2k2p^\prime 2k^\prime} |\mathcal{M}^{gg}_{gg}|^2(p,k,p^\prime,k^\prime) (2\pi)^4 \delta^{(4)}(p^\mu+k^\mu-p^{\prime \mu}-k^{\prime \mu}) \nonumber\\
        &\times \left[f_\vp f_\vk (1+f_{\vk^\prime})(1+f_{\vp^\prime})-f_{\vp^\prime} f_{\vk^\prime} (1+f_{\vp})(1+f_{\vk}) \right]\\
        \mathcal{C}^{1\leftrightarrow 2}[f](t,\vp) &= -\frac{(2\pi)^3}{8 p^2(N^2_c-1)} \int_0^\infty dp^\prime dk^\prime \left\{\gamma^g_{gg}(p;p^\prime,k^\prime) \delta^{(1)}(p-p^\prime-k^\prime) \left[f_{p\hat{\vp}}(1+f_{p^\prime \hat{\vp}})(1+f_{k^\prime \hat{\vp}})-f_{p^\prime \hat{\vp}} f_{k^\prime \hat{\vp}} (1+f_{p\hat{\vp}}) \right]\right.\nonumber\\
        &\left. -2\gamma^g_{gg}(p^\prime;p,k^\prime) \delta^{(1)}(p^\prime-p-k^\prime) \left[f_{p^\prime \hat{\vp}}(1+f_{p\hat{\vp}})(1+f_{k^\prime \hat{\vp}} )- f_{p\hat{\vp}} f_{k^\prime \hat{\vp}} (1+f_{p^\prime \hat{\vp}}) \right]\right\}\label{eq.app_c12},
    \end{align}
\end{subequations}
where we used the abbreviations $f_\vp=f(t,\vp)$, $p=|\vp|$, and $\hat{\vp}=\vp/p$ is the direction of collinear splitting with corresponding rate $\gamma^g_{gg}$. $|\mathcal{M}^{gg}_{gg}|^2$ is the $2\leftrightarrow 2$ scattering matrix element averaged over spin and color degrees of freedom. In vacuum, the corresponding expression \cite{Arnold:2002zm}
\begin{equation}
|\mathcal{M}^{gg}_{gg}|^2(p,k,p^\prime,k^\prime)=16(N^2_c-1) N^2_c g^4 \left(3-\frac{s_M t_M}{u^2_M} - \frac{s_M u_M}{t^2_M}-\frac{t_M u_M}{s^2_M} \right)    
\end{equation}
contains an infrared divergent term $(u_M-s_M)/t_M\sim 1/q^2$ (also in the $u_M$-channel) with $q$ the $t_M$-channel momentum transfer for a soft gluon exchange and $u_M,s_M,t_M$ denote the usual Mandelstam variables. This is regulated by inclusion of necessary physical interactions with the medium, where the leading thermal corrections are obtained as \cite{AbraaoYork:2014hbk,Kurkela:2018oqw} 
\begin{equation}
\label{eq.app_replacement}
\frac{u_M-s_M}{t_M} \to \frac{u_M-s_M}{t_M} \frac{q^2}{q^2+\xi^2_g m_D(t)^2}= \frac{\bar{u}_M-\bar{s}_M}{\bar{t}_M} \frac{\bar{q}^2}{\bar{q}^2+\xi^2_g A(t)\bar{m}^2_D}
\end{equation}
with $\xi_g = e^{5/6}/2$. The inclusion of screening effects however makes the scattering matrix element not invariant under rescaling as indicated due to different scaling of the Debye mass 
\begin{equation}
\label{eq.app_mD}
    m_D(t)^2 = 2g^2N_c\int\frac{d^3 \vp}{(2\pi)^3 p} f(t,\vp)=A(t)B(t)^{-2}\bar{m}^2_D\;.
\end{equation}
Moreover, the time-dependent Debye mass enters the scattering matrix element via Eq.~(\ref{eq.app_replacement}) in the denominator such that it can not be factored out thereby violating the assumption of overall scaling for the elastic collision kernel. The separation of variables in QCD kinetic theory can thus not generally be performed, as $m_D(t)$ leads to a mixing of time and (rescaled) momentum scales in the prescaling regime.

We now show that the inelastic collision kernel (\ref{eq.app_c12}) scales with $\mu_\alpha=3$ and $\mu_\beta=-1$ under prescaling in the non-expanding system~\eqref{eq.nonexp_prescaling}.
We consider only the first term since both have the same scaling behavior. Crucially, we need to know the scaling behavior of the splitting rate, which can be extracted from its two prevalent limiting regimes for soft gluon radition $z=p^\prime/p\ll 1$: the Bethe-Heitler (BH) limit in which interferences between successive scatterings are negligible and the Landau-Pomeranchuk-Migdal (LPM) limit in which successive scattering events by the medium interfere destructively. The rate in the respective limit reads
\begin{equation}
    \gamma^g_{gg}(p;p^\prime,k^\prime)|^{z\ll 1}_{\mathrm{BH}} \sim \frac{\hat{q}(\mu) p}{m_D(t)^2} |_{\mu = e m_D(t)},\qquad \gamma^g_{gg}(p;p^\prime,k^\prime)|^{z\ll 1}_{\mathrm{LPM}}  \sim \sqrt{\hat{q}(\mu) p}\;,
\end{equation}
where we only included those terms relevant for a scaling analysis with diffusion coefficient
\begin{equation}
    \hat{q}(\mu) \sim \log\frac{\mu^2}{2m_D(t)^2} \int \frac{d^3 \vp}{(2\pi)^3} f_\vp(1+f_\vp)\simeq  \log \frac{\mu^2}{2 A(t)B(t)^{-2} \bar{m}^2_D} A(t)^2 B(t)^{-3}\int \frac{d^3 \bar{\vp}}{(2\pi)^3} f_{S,\bar{\vp}} f_{S,\bar{\vp}}
\end{equation}
in this overoccupied scenario. For $\hat{q}(\mu)|_{\mu=em_D}$ in the BH limit, we thus have $\hat{q}(\mu)=A(t)^2 B(t)^{-3} \bar{\hat{q}}$ and
\begin{equation}
    \gamma^g_{gg}(p;p^\prime,k^\prime)|^{z\ll 1}_{\mathrm{BH}}\sim \frac{A(t)^2 B(t)^{-3} \bar{\hat{q}} B(t)^{-1} \bar{p}}{A(t) B(t)^{-2} \bar{m}^2_D} \sim A(t) B(t)^{-2}\bar{\gamma}^g_{gg}|^{z\ll 1}_{\mathrm{BH}}\;.
\end{equation}
For the LPM limit, $\mu$ is to next-to-leading-logarithmic order given self-consistently via 
\begin{equation}
    \mu^2\sim \sqrt{\hat{q}(\mu)p}\sim \sqrt{A(t)^2 B(t)^{-3}\log\left( \frac{\mu^2}{2A(t)B(t)^{-2}\bar{m}^2_D}\right) B(t)^{-1} } \sim A(t) B(t)^{-2} \sqrt{\log\left( \frac{\mu^2}{2A(t)B(t)^{-2}\bar{m}^2_D}\right)}\;,
\end{equation}
which shows that $\mu$ scales self-consistently as $\mu^2 \sim A(t) B(t)^{-2}$ such that the time dependence drops out of $\log(\mu^2/(2m^2_D))$. Accordingly, for the LPM limit we thus find again
\begin{equation}
    \gamma^g_{gg}(p;p^\prime,k^\prime)|^{z\ll 1}_{\mathrm{LPM}}  \sim \sqrt{\hat{q}(\mu) p} \sim \sqrt{A(t)^2 B(t)^{-3}\bar{\hat{q}}(\mu) B(t)^{-1} \bar{p}} \sim A(t) B(t)^{-2} \bar{\gamma}^g_{gg} |^{z\ll 1}_{\mathrm{LPM}}\;.
\end{equation}
The rate therefore scales like the Debye mass in both limits and we will adopt this scaling for the complete rate $\gamma^g_{gg} \sim A(t) B(t)^{-2} \bar{\gamma}^g_{gg}$. This leads to the overall scaling prediction
\begin{align}
    \mathcal{C}^{1\leftrightarrow 2}[f](t,\vp) &\sim \frac{1}{B(t)^{-2} \bar{p}^2} \int_0^\infty B(t)^{-2} d\bar{p}^\prime d\bar{k}^\prime  A(t) B(t)^{-2} \bar{\gamma}^g_{gg}(\bar{p};\bar{p}^\prime,\bar{k}^\prime) B(t) \delta^{(1)}(\bar{p}-\bar{p}^\prime-\bar{k}^\prime) \nonumber\\
    &\qquad \qquad \qquad \times A(t)^2\left[f_{S,\bar{p}\bar{\hat{\vp}}}(f_{S,\bar{p}^\prime \bar{\hat{\vp}}}+f_{\bar{k}^\prime \bar{\hat{\vp}}}) - f_{S,\bar{p}^\prime \bar{\hat{\vp}}} f_{S,\bar{k}^\prime \bar{\hat{\vp}}} \right]\\
    &\sim A(t)^3 B(t)^{-1} \mathcal{C}^{1\leftrightarrow 2}[f_S](\bar{\vp})\label{eq.app_c12_scaling}\;,
\end{align}
such that we can identify $\mu_\alpha=3$ and $\mu_\beta=-1$ as anticipated. $\mathcal{C}^{1\leftrightarrow 2}$ will therefore lead to the direct energy cascade fixed point $(\alpha_\infty,\beta_\infty)=(-4/7,-1/7)$.\\
The scaling analysis for the elastic collision kernel in the absence of the Debye mass has been performed in \cite{Berges:2013fga} and can simply be generalized to the prescaling case with again $\mu_\alpha=3$ and $\mu_\beta=-1$. This scaling analysis however needs to be augmented by the inclusion of the Debye mass as we discussed above, which leads to an effective regulation of the divergent soft contributions to the elastic collision kernel and prevents one from extracting an overall scaling behavior thereof.

\section{Fokker-Planck approximation}
We assume the gluons to interact via elastic small-angle scatterings such that the collision kernel takes a FP form \cite{Mueller:1999pi,Schlichting:2019abc,Berges:2020fwq,Brewer:2022vkq}
\begin{align}
    \mathcal{C}^{\mathrm{FP}}(t,\vp) &\sim \log \frac{\langle p \rangle(t)}{m_D(t)}  \left[\int \frac{d^3 \vk}{(2\pi)^3} f_\vk (1+f_\vk) \; \partial^2_p f_\vp + \int \frac{d^3 \vk}{(2\pi)^3 k} 2 f_\vk \;\vec{\nabla}_{\vp}\cdot \left( \frac{\vp}{p} (1+f_\vp)f_\vp\right) \right]\\
    &\equiv \log \frac{\langle p \rangle(t)}{m_D(t)} \tilde{C}^{\mathrm{FP}}[f](t,\vp),
\end{align}
where we highlighted the contributions relevant for a scaling analysis and here $\langle \dots \rangle \equiv \int \frac{d^3 \vp}{(2\pi)^3} \dots f(t,\vp)/\int \frac{d^3 \vp}{(2\pi)^3} f(t,\vp)$. The advantage of the FP approximation for a prescaling analysis becomes apparent here, as the contribution due to the Debye mass in the elastic QCD scattering matrix element is simply factorized into a logarithm of the characteristic UV and IR scale.
Plugging in the prescaling ansatz, we then find directly for the overoccupied system that
\begin{align}
\label{eq.app_cFP_rescaling}
    \mathcal{C}^{\mathrm{FP}}[f](t,\vp) &= \log\left[  \frac{\langle \bar{p} \rangle}{A(t)^{\tfrac{1}{2}} \bar{m}_D} \right] A(t)^3 B(t)^{-1} \tilde{C}^{\mathrm{FP}}[f_S](\bar{\vp}). 
\end{align}
\section{Breaking of scaling}
\begin{figure}[t!]
    \centering
    \includegraphics[width=0.55\linewidth]{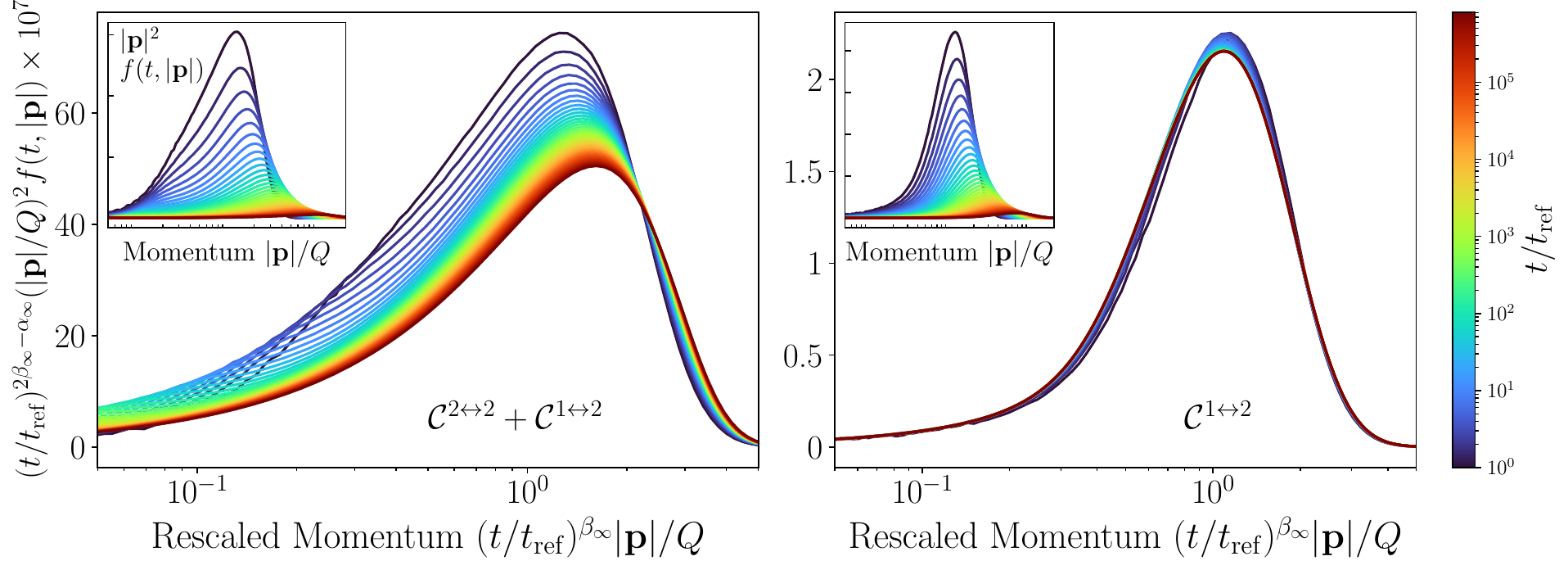}
    \includegraphics[width=0.44\linewidth]{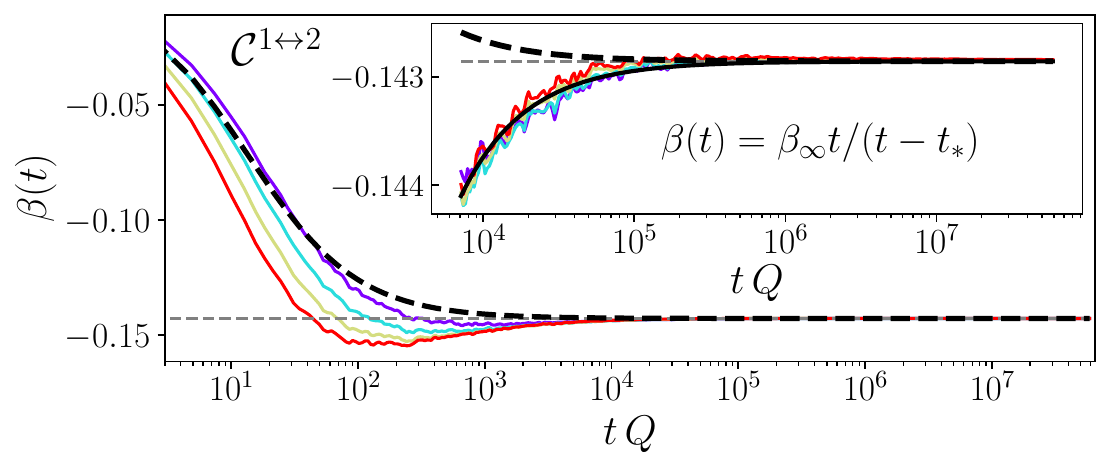}    
    \caption{(Left) Gluon distribution function $|\vp|^2 f(t,|\vp|)$ from kinetic theory simulation with (left) elastic and inelastic and (right) only inelastic  processes rescaled with fixed point exponents $(\alpha_\infty,\beta_\infty)=(-4/7,-1/7)$, with $t_{\mathrm{ref}}\,Q\sim 66$. The unrescaled distributions are shown in the insets. (Right) Comparison of prescaling exponents from inelastic-only simulations to analytical expectation~\eqref{eq.nonexp_general_sol} of the main text. We determine $t_*=t_0(1-\beta_\infty/\beta_0)$ from the data at $t_0\,Q\sim 2$ (dashed black line) and at $t_0\,Q\sim 10^4$ (solid black line). Different colors represent prescaling exponents extracted from different integral moments, see Eq.~\eqref{eq.nonexp_exponents_moments}, where we depict $\beta_{m=0,1,2,3}$ with colors $(0,1,2,3)\to$(violet, turquoise, yellow, red).}
    \label{fig.app_nonexp_fresc}
\end{figure}
The breaking of scaling by the Debye mass in the elastic collision kernel discussed in the main text is visualized in the left and middle panel of Fig.~\ref{fig.app_nonexp_fresc}, where we show $|\vp|^2 f(t,|\vp|)$ rescaled with fixed point exponents. The rescaling for simulations with only inelastic scatterings (middle) shows a very clear collapse of all curves from shortly after initialization over an evolution of six orders of magnitude, but a spread cannot be removed for all times even by time-dependent rescalings if one includes elastic scatterings (top left) due to the presence of the Debye mass. In the right panel we extract prescaling exponents for only inelastic scatterings ${\cal C}^{1\leftrightarrow 2}$ from different moments of the distribution function (see Eq.~\eqref{eq.nonexp_exponents_moments}). The resulting approach of $\beta(t)$ to the fixed point value $\beta_\infty=-1/7$ (dashed gray line) is demonstrated to be well-described over more than seven orders of magnitude in time by Eq.~\eqref{eq.nonexp_general_sol} of the main text (dashed black line) with $\mu_\alpha=3$ and $\mu_\beta=-1$, see Eq.~\eqref{eq.app_c12_scaling}, even at surprisingly early times shortly after initialization. In the inset, we show that this solution initialized at a very early time $t_{0} \, Q\sim 2$ with $t_{*}$ obtained from $\beta(t_{0}) \equiv \beta_{0}$ captures the evolution at intermediate times $t\,Q\sim 10^4$ only qualitatively. If we obtain $t_{*}$ in the same way at a later time $t_0\, Q\sim 10^4$, Eq.~\eqref{eq.nonexp_general_sol} describes the late-time evolution quantitatively well as given by the solid black line.
 
\section{Derivation of Eq. (8)}
Here we give a brief account of how one derives Eq.~\eqref{eq.nonexp_Bode_FP} and Eq.~\eqref{eq.nonexp_betaode_FP} of the main text. As we demonstrated above, the FP collision kernel does not scale homogeneously under the prescaling ansatz and the corresponding prescaling solutions are therefore not captured by Eq.~\eqref{eq.nonexp_general_sol} of the main text. The $t$- and $\bar{\vp}$-dependent contributions of the FP collision kernel in Eq.~\eqref{eq.fp_collision} of the main text however factorize
\begin{align}
    \partial_t f(t,\vp) &=A(t) \left[\frac{\dot{A}(t)}{A(t)}+ \frac{\dot{B}(t)}{B(t)} \bar{\vp} \cdot \partial_{\bar{\vp}}\right]f_S(\bar{\vp}) \\
    &\stackrel{A=B^\sigma}{=}B(t)^{\sigma-1} \dot{B}(t) \left[ \sigma +\bar{\vp}\cdot \partial_{\bar{\vp}}\right]f_S(\bar{\vp})\\
    &=\log\left[  \frac{\langle \bar{p} \rangle}{B(t)^{\tfrac{\sigma}{2}} \bar{m}_D} \right] B(t)^{3\sigma-1} \tilde{C}^{\mathrm{FP}}[f_S](\bar{\vp}),
\end{align}
which is the necessary property to perform separation of variables. Similarly to the derivation of Eq.~\eqref{eq.nonexp_Bode_general} one can now separate variables in the corresponding Boltzmann equation
\begin{align}
\frac{ B(t)^{-2\sigma} \dot{B}(t)}{\log\left[  \frac{\langle \bar{p} \rangle}{B(t)^{\tfrac{\sigma}{2}} \bar{m}_D} \right]} &=D_2 = \frac{\tilde{C}^{\mathrm{FP}}[f_S](\bar{\vp})}{[\sigma+\bar{\vp}\cdot \partial_{\bar{\vp}}]f_S(\bar{\vp})}
\end{align}
with $D_2$ the associated separation of variables constant. Reorganizing the LHS
\begin{equation}
\label{eq.app_fp_Bode}
    \partial_t B(t) = D_2 B(t)^{2\sigma} \left[\log\left(\frac{\langle \bar{p}\rangle}{\bar{m}_D}  \right)- \frac{\sigma}{2} \log B(t)\right],
\end{equation}
one can recognize Eq.~\eqref{eq.nonexp_Bode_FP} of the main text with $D_2$ fixed for convenience at time $t_0$ where $B(t_0)=1$ such that
\begin{align}
    \partial_t B(t) |_{t_0} &=\frac{\beta_0}{t_0} \\
    &=D_2 \log\left(\frac{\langle \bar{p}\rangle}{\bar{m}_D}  \right)\\
    \Leftrightarrow \quad D_2 &= \frac{\beta_0}{t_0 \log\left(\frac{\langle \bar{p}\rangle}{\bar{m}_D}  \right)}. \label{eq.app_nonexp_Bode_FP}
\end{align}
We note that this is an intricate self-consistent equation as $\log[\langle \bar{p}\rangle/\bar{m}_D]$ depends implicitly on $D_2$ via $f_S$. We avoid the need to solve it as we utilize our simulations to obtain $\langle\bar{p}\rangle/\bar{m}_D$ as a function of $\beta_{0}/t_{0}$.

We now rewrite Eq.~\eqref{eq.app_fp_Bode} in terms of $\beta(t)$
\begin{equation}
    \frac{\beta(t)}{t} = D_2 e^{(2\sigma-1) \int_{t_0}^t dt^\prime \frac{\beta(t^\prime)}{t^\prime}} \left[\log \frac{\langle \bar{p}\rangle}{\bar{m}_D}-\frac{\sigma}{2} \int_{t_0}^t dt^\prime \frac{\beta(t^\prime)}{t^\prime} \right]
\end{equation}
and take two temporal derivatives to obtain Eq.~\eqref{eq.nonexp_betaode_FP}. The first derivative yields
\begin{align}
    \partial_t\beta(t) &= \frac{\beta(t)}{t} + (2\sigma-1) \frac{\beta(t)^2}{t} -2D_2 \beta(t) e^{(2\sigma-1) \int_{t_0}^t dt^\prime \frac{\beta(t^\prime)}{t^\prime}}\\
    \dot{\beta}(t_0) &= \beta(t_0) \left[\frac{1}{t_0}+(2\sigma-1) \frac{\beta(t_0)}{t_0}-2D_2 \right],
\end{align}
where for the second equation we evaluated the first equation at $t_0$ to obtain constraints on the initial data. A second temporal derivative then reproduces Eq.~\eqref{eq.nonexp_betaode_FP} of the main text, where we can use the first order equation $\dot{\beta}$ to get rid of the $D_2$ dependence
\begin{equation}
\label{eq.app_nonexp_betaode_FP}
    \frac{\ddot{\beta}(t)}{\beta(t)} =\frac{\dot{\beta}(t)^2}{\beta(t)^2}+\frac{14 \dot{\beta}}{t}- \frac{(7\beta(t)+1)^2}{t^2}.
\end{equation}
This equation is also subtle, as its second order character stays in contrast with the number of parameters needed to solve Eq.~\eqref{eq.app_nonexp_Bode_FP}, which requires specifying only $\beta_{0}$ at~$t_{0}$. Indeed, there is a nontrivial constraint on initial data for Eq.~\eqref{eq.app_nonexp_betaode_FP} directly following from a derivative of Eq.~\eqref{eq.app_nonexp_Bode_FP}:
$\dot{\beta}(t_{0}) \equiv \dot{\beta}_{0} =\tfrac{\beta_{0}}{t_{0}}[ 1 - \beta_{0}/\beta_\infty - 1/ \log{\sqrt{\langle\bar{p}\rangle/\bar{m}_D}}]$.
Its complexity arises from the dependence of $\langle\bar{p}\rangle/\bar{m}_D$ on $\beta_{0}$ and $t_{0}$. For the solution (dashed black line) in Fig. \ref{fig.nonexp_prescaling_FPpred} of the main text, the initial conditions at $t_0$ for $\dot{\beta}_{0}$ are determined by the constraint and $\beta_{0}$ (as well as $\log\sqrt{\langle \bar{p}\rangle/\bar{m}_D}$) extracted from the data to consistently compare to the QCD kinetic theory simulations.


\section{Large order behavior of transseries}

\begin{figure}[h!]
    \centering
    \includegraphics[width=0.6\columnwidth]{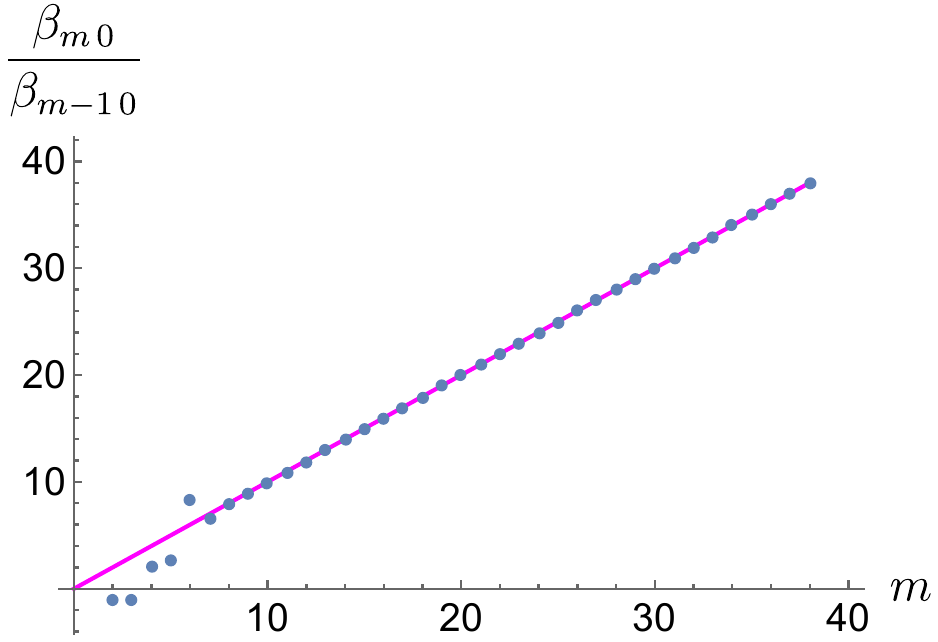}    \caption{Ratio test indication that the series~({\color{blue}9a}) has a vanishing radius of convergence}
    \label{fig.largeorderbeh}
\end{figure}

We iteratively solved for the coefficients in Eq.~\eqref{eq.lateansatz} of the main text in Mathematica. This allowed us to generate exactly within several hours the lowest 38 orders of the expansion in inverse powers of $\log{(Q\,t)}$. In Fig.~\ref{fig.largeorderbeh} we show that the generated coefficients $\beta_{m,0} \sim m!$ for $m$ already as small as 10. Furthermore, we also found that the leading late time coefficient at each order obey $\beta_{m , m-1} = (-1)^{m} \frac{1}{7}$, i.e. they form a geometric series allowing to simply resum their contribution.

\end{appendix}

\end{document}